\documentclass{svproc}

\renewcommand{\bfseries}{\fontseries{b}\selectfont}

\usepackage{graphicx}%
\usepackage{multirow}%
\usepackage{amsmath,amssymb,amsfonts}%
\usepackage{mathrsfs}%
\usepackage[title]{appendix}%
\usepackage{xcolor}%
\usepackage{textcomp}%
\usepackage{manyfoot}%
\usepackage{booktabs}%
\usepackage{algorithm}%
\usepackage{algorithmicx}%
\usepackage{algpseudocode}%
\usepackage{listings}%
\usepackage{caption} 
\usepackage{subcaption} 

\usepackage{url}

\def\orcidID#1{\unskip$^{[#1]}$} 

\begin{document}
\mainmatter              
\title{Dynamic Link and Flow Prediction \\in 
Bank Transfer Networks}
\titlerunning{Dynamic Link and Flow...}  
%
\author{Shu Takahashi\inst{1} \and Kento Yamamoto\inst{1} \and Shumpei Kobayashi\inst{1} \and Ryoma Kondo\inst{1,2}  \and Ryohei Hisano\inst{1,2}\orcidID{https://orcid.org/0000-0002-4452-0302}}
\authorrunning{Shu Takahashi et al.} 

\institute{Graduate School of Information Science and Technology, the University of Tokyo, 7-3-1 Hongo, Bunkyo-ku, Tokyo, 113-8656, Japan\\
\email{hisanor@g.ecc.u-tokyo.ac.jp}
\and
The Canon Institute for Global Studies,
11th Floor, ShinMarunouchi Building, 5-1 Marunouchi 1-chome, Chiyoda-ku, Tokyo 100-6511, Japan}

\maketitle              

\begin{abstract}

The prediction of both the existence and weight of network links at future time points is essential as complex networks evolve over time. Traditional methods, such as vector autoregression and factor models, have been applied to small, dense networks, but become computationally impractical for large-scale, sparse, and complex networks. Some machine learning models address dynamic link prediction, but few address the simultaneous prediction of both link presence and weight. Therefore, we introduce a novel model that dynamically predicts link presence and weight by dividing the task into two sub-tasks: predicting remittance ratios and forecasting the total remittance volume. We use a self-attention mechanism that combines temporal-topological neighborhood features to predict remittance ratios and use a separate model to forecast the total remittance volume. We achieve the final prediction by multiplying the outputs of these models. We validated our approach using two real-world datasets: a cryptocurrency network and bank transfer network.

\keywords{Temporal Networks, Deep Learning, Bank Transfer Networks}

\end{abstract}

\section{Introduction}\label{sec1}
In recent years, the concept of dynamic networks has become increasingly popular for elucidating the evolving nature of relationships in diverse systems, encompassing social, biological, and physical domains. The burgeoning interest in modeling and analyzing these networks is attributable to their capacity to offer profound insights into the dynamic intricacies of these systems. In \cite{Peel2015}, the authors made significant contributions to this field by investigating the dynamics of social relationships, using data derived from Bluetooth scans of mobile phones, as initially explored in \cite{Eagle2006}. In the realm of neuroscience, in \cite{Patil2021}, the authors highlighted the significance of temporal connections between major brain networks in the context of creative cognition, drawing on functional magnetic resonance imaging data. Furthermore, in the domain of public transportation networks, in \cite{Wu2019}, the authors developed an innovative graph neural network framework that effectively reveals hidden spatial dependencies, which marks a noteworthy advancement in this area.

Dynamic link prediction is a pivotal task in the realm of dynamic networks~\cite{Trivedi2018,Wang2021,Poursafaei2022,Zhang2022}. In \cite{Xu2020}, the authors presented  the temporal graph attention (TGAT) layer, which leverages a self-attention mechanism paired with a time encoding technique based on Bochner’s theorem. This approach efficiently amalgamates temporal-topological neighborhood features, thereby enabling  the model to generate node embeddings from time-dependent functions, which makes them appropriate for dynamic link prediction. By contrast, in \cite{Wang2021}, the authors used temporal random walks to intuitively depict temporal networks. Their innovative approach captures network dynamics and incorporates an astute anonymization strategy for link prediction. Although these contributions are noteworthy, recently, in \cite{Poursafaei2022}, the authors highlighted a common oversight: many dynamic networks have persistent links, and several prediction models neglect this simple reality. Surprisingly, numerous models do not outperform the elementary prediction method of ``EdgeBank,'' where the prediction merely relies on links that persisted over a previous time window.

Relationships in a dynamic network can also be described by continuous weights, as seen in the Bitcoin alpha network~\cite{Kumar2016}, where each weight represents the trust between two nodes. Even more intriguing, these weights can sometimes directly indicate the flow of substances or information within a network, as observed in global trade networks~\cite{Brauning2020}, interbank networks~\cite{Giraitis2016,Linardi2020,Zhang2022}, and bank transfer networks. In such cases, the concentration of substance or money in a network's upper layers, as highlighted by concepts such as the Helmholtz--Hodge potential~\cite{Kichikawa2019} or trophic analysis~\cite{MacKay2020}, might suggest forthcoming fund transfers to the lower layers. Additionally, a growing number of tightly interlinked components in an interbank network, particularly when accounting for their edge weights, could indicate heightened interconnectedness. This pattern may also serve as an early warning of rising systemic risks~\cite{Zhang2022,Denbee2021}.

A specific area of interest that focuses on predicting link weights in machine learning is the field of flow prediction, where the objective is to predict the flow (or edge weights) of established edges. Distinct from traditional link prediction, the emphasis in this area is on already known edges and the aim is to interpolate their weights. The seminal work in \cite{Jia2019} addressed flow prediction. The authors proposed a graph-based semi-supervised learning method to predict edge flows while taking flow conservation constraints into account. A more contemporary contribution is \cite{Silva2021}, which offers an advanced optimization framework that blends bilevel optimization with deep learning to predict missing edge flows using partial data and edge attributes. These studies primarily engage with traffic and power transmission networks, whereas in \cite{Ceylan2021}, the authors made significant strides in the context of bank transaction networks. Their approach refines the gradient model (i.e., \cite{Kichikawa2019}) to a gated variant, introduces a multi-scale, outlier-resistant loss function, and achieves superior results compared with leading models. Although these studies are interesting in their own right, a common limitation is their inability to predict future networks.


The challenge of dynamic link flow prediction stands in stark contrast to the scope of existing literature. In previous research, such as studies forecasting world trade networks~\cite{Dubovik2022}, researchers typically concentrated on small and densely connected networks. The clear delineation of existing edges in these networks allows for the application of vector autoregressive models ~\cite{Bussiere2009}, a dynamic latent space model~\cite{Ward2013}, and dynamic factor models~\cite{Brauning2020} to model their dynamics. In this context, these studies are similar to flow prediction scenarios in which network connections are predetermined. By contrast, our focus is on modeling complex networks~\cite{Newman2010} characterized by sparsity and scale-free behavior that makes it important to predict both the existence of edges and their weights. Given their ubiquity in real-world contexts, there is a significant impetus to develop methodologies for such complex networks.

\begin{figure}[!ht]
 \centering
 \includegraphics[scale=0.5]{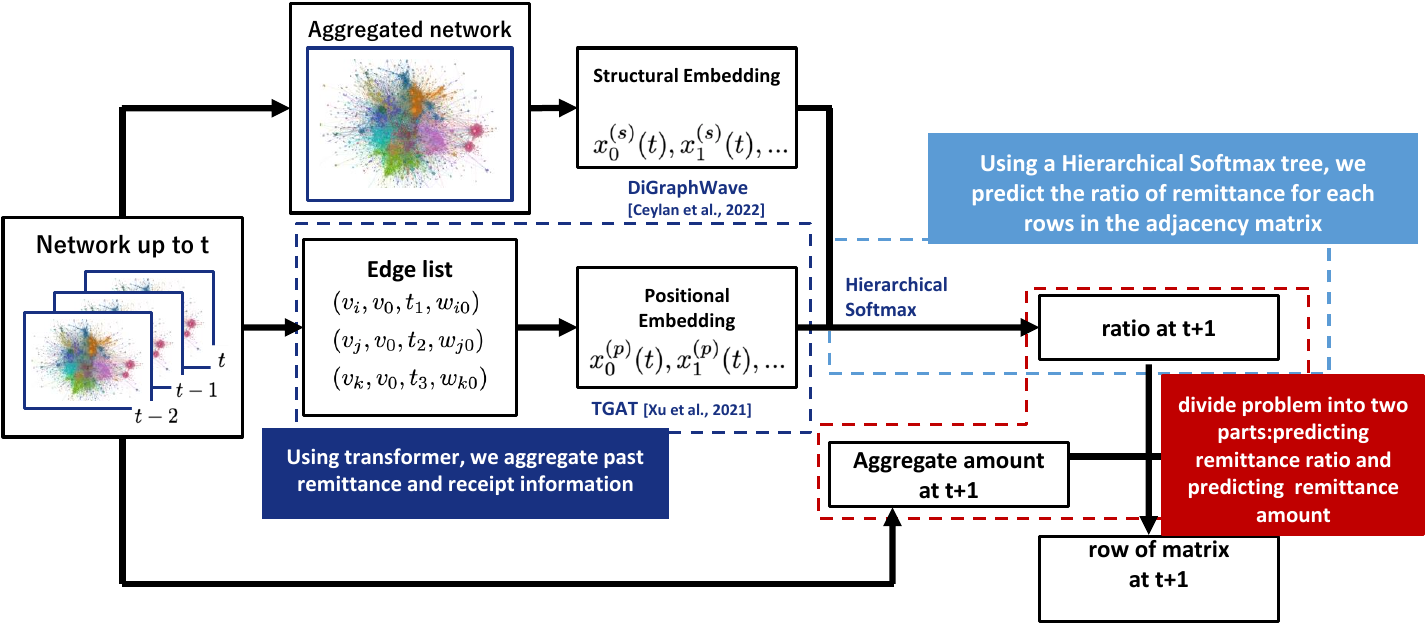}
 \caption{Description}
 \label{fig:model_concept}
\end{figure}

In this study, we introduce an innovative model designed to concurrently predict dynamic links and flow in networks. This model addresses the challenge of flow prediction by estimating the distribution of flows that a node directs toward others and multiplying this estimate by the node's projected total transmission in the subsequent time period. Our approach integrates four primary components as depicted in Fig.\ref{fig:model_concept}: (1) structural embedding, which leverages historical network data to position nodes according to their structural roles, thereby effectively capturing their enduring characteristics within the network~\cite{Ceylan2022}; (2) transfer activity embedding, which is inspired by TGAT~\cite{Xu2020} and focuses on recent transfer activities to reflect the current operational status of nodes; (3) a shallow and wide hierarchical softmax tree, which uses a combination of structural and output features with top-down clustering to mitigate error propagation in tree-based methods, thereby enhancing the accuracy of predicting flow proportion distributions among network nodes~\cite{Khandagale2020}; and (4) a projection mechanism for estimating the total amount a node will transmit in the next time step. Together, these elements enable our model to proficiently predict dynamic link flows, thereby providing a robust framework for analyzing and understanding complex network behaviors.

We thoroughly evaluated our dynamic link flow prediction model using proprietary bank transaction data from a major Japanese bank, in addition to publicly available cryptocurrency transfer data. We compared our model with previous models for both the remittance ratio and actual weights, with additional analysis excluding persistent edges. Our proposed model demonstrated superior performance in terms of ratio prediction, whereas the results for actual weight prediction were mixed. The challenge in predicting weights was particularly evident for nodes with sporadic activity. Additionally, our model performed well in predicting link formation and dissolution.

\section{Data}\label{data}

We provide a brief overview of the datasets used. The bank transfer data were supplied by a Japanese bank. The data focus on the top companies with at least 1,000 transactions during the two-year period from April 2019 to February 2021. The cryptocurrency dataset is based on open-source data from \cite{Beres2020}. Given that many accounts appear only once or twice in the dataset, we concentrate on the top accounts that consistently appear during the period from April 2018 to April 2020. The reformatted version of the latter dataset is available in our repository\footnote{\url{https://github.com/hisanor013/DynamicLinkFlow}}.

\begin{table}[h]
\centering
\caption{Comparison of network statistics for the Ethereum and Bank datasets.}
\label{tab:network_stats}
\begin{tabular}{@{}lcc@{}}
\toprule
\textbf{Metric}                  & \textbf{Ethereum}   & \textbf{Bank}      \\ \midrule
\textbf{Number of snapshots}      & 25                 & 23                 \\
\textbf{Number of nodes}          & 476                & 14,976             \\
\textbf{Number of edges}          & 20,612             & 11,878,442         \\
\textbf{Average sparsity}         & 0.0073             & 0.0046             \\
\textbf{Average edge persistence } & 0.4560             & 0.4991             \\ \bottomrule
\end{tabular}
\end{table}
\begin{figure}[th!]
    \centering
    \begin{minipage}[b]{0.4\textwidth}
        \centering
        \includegraphics[width=\textwidth]{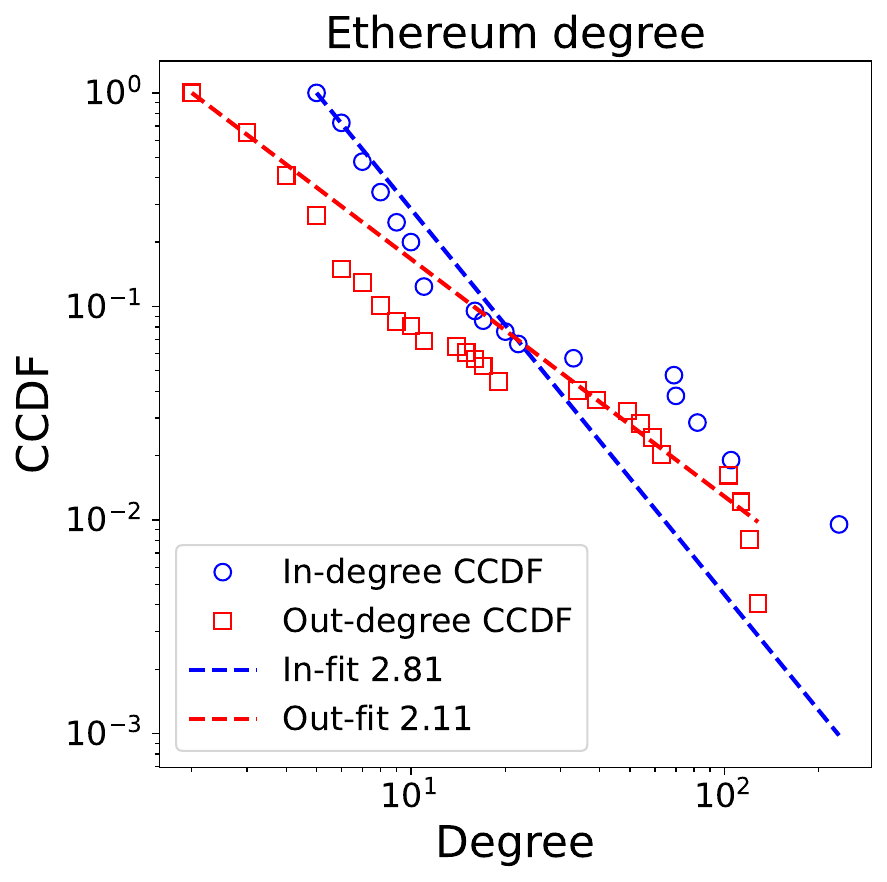}
    \end{minipage}
    \begin{minipage}[b]{0.4\textwidth}
        \centering
        \includegraphics[width=\textwidth]{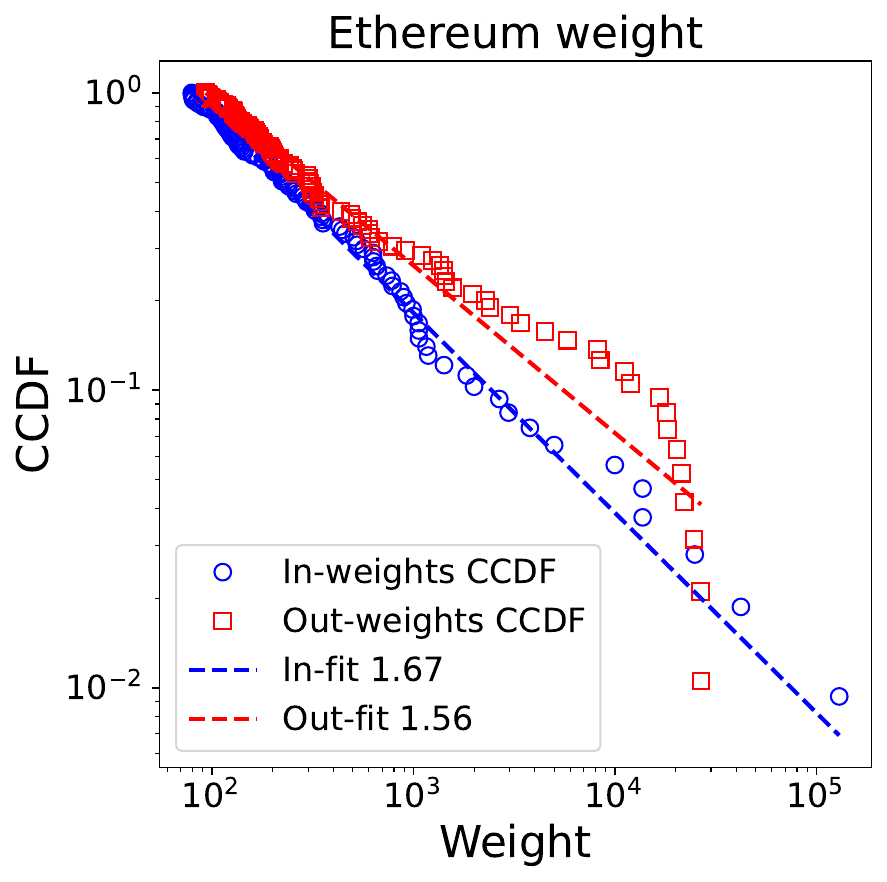}
    \end{minipage}
    \caption{Complementary cumulative distribution of both degrees and weights for the Ethereum dataset.}
    \label{cdf:eth}
\end{figure}
\begin{figure}[thp!]
    \centering
    \begin{minipage}[b]{0.4\textwidth}
        \centering
        \includegraphics[width=\textwidth]{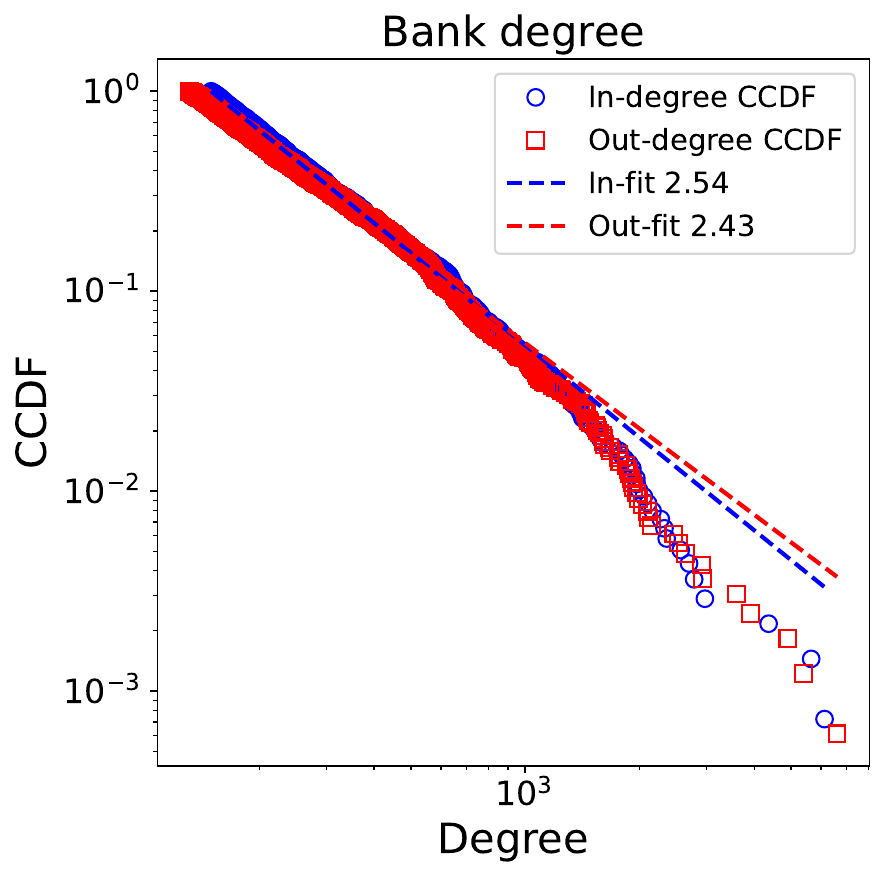}
    \end{minipage}
    \begin{minipage}[b]{0.4\textwidth}
        \centering
        \includegraphics[width=\textwidth]{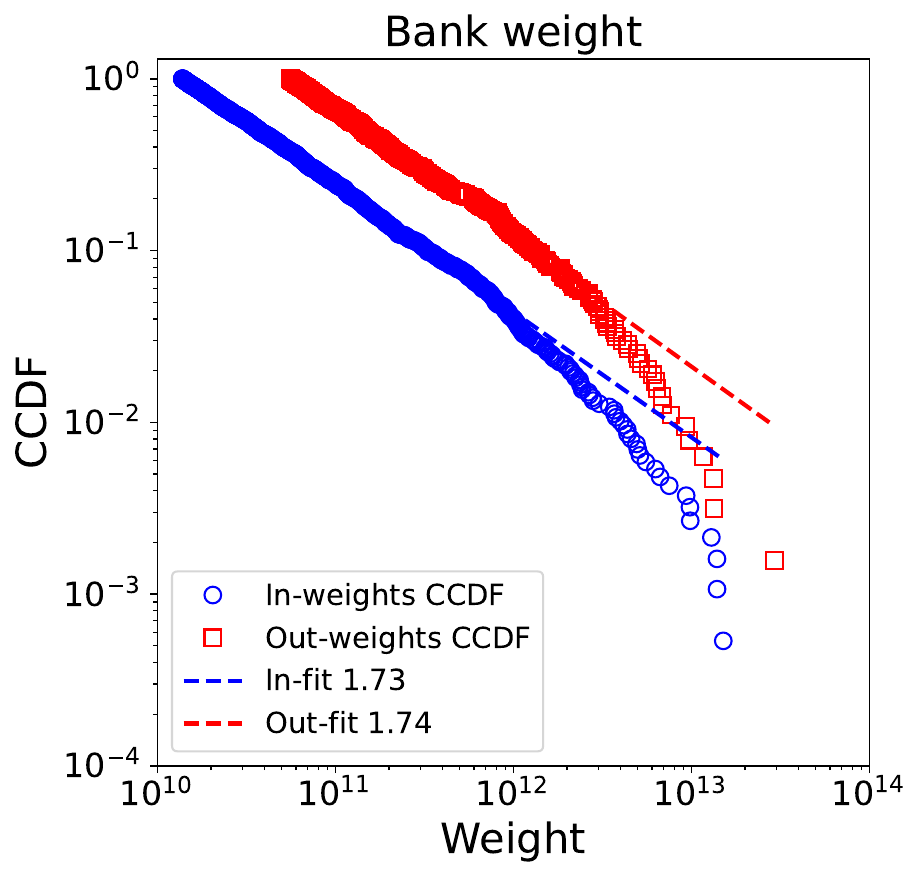}
    \end{minipage}
    \caption{Complementary cumulative distribution of both degrees and weights for the Bank dataset.}
    \label{cdf:bank}
\end{figure}

First, we calculate key statistics to understand the structure and dynamics of the temporal network. We determine the number of snapshots (distinct time points), nodes (unique entities), and edges (connections between nodes) to provide a basic overview of the network. Additionally, we compute the average sparsity, which measures how densely connected the network is over time by comparing the actual number of edges to the maximum possible connections in a fully connected network. We also calculate the average edge persistence, which indicates how consistently connections between nodes reappear across different snapshots. To calculate this, we group edges by their unique source-target pairs and count how many times each edge appears across all snapshots, then average these counts and normalize them by the total number of snapshots. This helps us to assess the stability of relationships within the network over time. The result is summarized in Table~\ref{tab:network_stats}.

Moreover, we analyze the complementary cumulative distribution of both the degrees and weights of the aggregated network for both datasets, as shown in Figs.\ref{cdf:eth} and \ref{cdf:bank}. Although our focus on the most frequent accounts involved in remittances prevents us from observing a complete power-law distribution, as summarized in \cite{Hisano2024}, we still identify a heavy-tailed distribution in the upper tail. Additionally, in the figures, we provide an estimate of the power-law distribution using the method in \cite{Clauset2009,Alstott2014}.

\section{Model}\label{data}
Dynamic link flow prediction entails forecasting the future weighted adjacency matrix $A(t)$ based on data from previous weighted adjacency matrices $A(1), \ldots, A(t-1)$. From $A(t)$, we define the vector $w(t) \in \mathbb{R}_{\geq 0}^{N}$, which represents the total remittance amount for each node (i.e., bank account), and the matrix $R(t) \in [0, 1]^{N \times N}$, which indicates the remittance ratios, as follows:

\begin{equation}
w_{i}(t) = \sum_{j=1}^{N} A_{ij}(t), \: R_{ij}(t) = \frac{A_{ij}(t)}{w_{i}(t)}.
\end{equation}


Our model predicts $w(t)$ and $R(t)$ independently rather than directly predicting the adjacency matrix as shown in Fig.~\ref{fig:model_concept}. In all the datasets used in this paper, $t$ is set on a monthly basis.

We predict $R(t)$ by aggregating information from recent transactions in past networks. To effectively gather information from neighboring nodes over time, we use a deep learning architecture similar to TGAT~\cite{Xu2020}, which uses a self-attention mechanism and temporal encoding. Let $X_{i} \in \mathbb{R}^{d}$ represent the features of node $i$, $\text{Ne}(i; t)$ denote the neighboring nodes of $i$ at time $t$, $M$ be the number of elements in the concatenated feature, and $\Phi(t) := \cos(tw + b)$ be the temporal encoding function. TGAT resembles basic graph convolutional networks (GCNs)~\cite{Bishop2024} in that it aggregates information from neighboring nodes; however, it differs in how it transforms this information and incorporates temporal information as an additional feature. Similarly to GCNs, the first layer transforms the node feature vector into a hidden state. The second layer further refines this information by considering neighboring nodes, the timing of interactions, and edge features, and applies a self-attention mechanism and nonlinear activation to transform the data. Finally, we process the embeddings to the remittance ratio using hierarchical softmax. The underlying operations can be expressed succinctly as follows:
\begin{alignat}{2} 
Z_{i}^{1}(t) &= \sigma(X_{i}), & \\
H_{i}(t) &= \left[ Z_{i}^{1}(t); \Phi(0); Z_{j_{1}} ; E_{ij_{1}}(t); \Phi(t-t_{j_{1}}), \dots, \right], & \\ 
Q(t) &= [H_{i}(t)]_{0}W_{Q}, K(t) = [H_{i}(t)]_{1:M} W_{K}, V(t) = [H_{i}(t)]_{1:M}W_{V}, & \\ 
\bar{Z}_{i}^{2}(t) &= \mathrm{Softmax} \left( \frac{Q(t)K(t)^T}{\sqrt{d}} \right)V(t), & \\ 
Z_{i}^{2}(t) &= \sigma \left( \left[\bar{Z}_{i}^{2}(t); X_{i}\right]W_{0} + b_{0} \right) W_{1} + b_{1}, &  \\
R_{i}(t) &= \text{HSoftmax}(Z_{i}^{2}(t)), &
\end{alignat}
\noindent where $\sigma$ denotes the nonlinear activation function, $j_{1}, \dots \in \text{Ne}(v_{i}; t)$, ``;'' represents concatenation, $d$ is the dimensionality of the key and query vectors, and $\text{HSoftmax}$ denotes hierarchical softmax.

We derive the node features (i.e., $X_{i}(t)$)  using DiGraphWave \cite{Ceylan2022}, which computes structural node embeddings. To prevent data leakage, we restrict the use of network information to the first three months of the dataset. For the edge features (i.e., $E_{ij}$), we include the logarithm of the transfer amount, the transfer ratio, the sender's ratio, the recipient's ratio, and the ratio calculated across the entire adjacency matrix.

The choice of a hierarchical softmax function is motivated by the observation that, when handling a large number of classes, applying a shallow and wide hierarchical softmax function can significantly improve performance~\cite{Khandagale2020}. Therefore, rather than using a flat softmax function, we implement a shallow and wide hierarchical softmax tree with a maximum depth of 3. We construct this hierarchy using the $K$-means algorithm applied to the combination of the aforementioned structural embeddings and output features as in \cite{Khandagale2020}.

In many GCN-style models, as the network size increases, neighborhood sampling becomes critical for computational efficiency, and TGAT is no exception. To address this challenge, we modify the original implementation by sorting transactions based on recency and weight, and then select the top 100 from this sorted list. Moreover, to mitigate the edge persistence effect, we model the probability by combining the average probability from all past occurrences with our prediction, using a mixing ratio of 8:2. This approach encapsulates our method for rate prediction.

For total remittance prediction, we use a gradient boosting model, using each node's past transactions as input features. We derive the features in the prediction model  from various transactional aspects of the node over the previous three months. Specifically, the model includes the total amount of money sent and received by the node during this period as basic transaction features. Additionally, it incorporates the difference between sent and received amounts for each month, providing insights into the net transaction flow. The model also captures month-to-month changes in both the sent and received amounts, allowing it to account for fluctuations in transaction patterns over time. By combining these features, the model aims to predict future transaction volumes more accurately. We set the learning rate to 0.0001, with 20,000 estimators, and limit the maximum model depth to 4.

\section{Results}\label{data}
We evaluated the predictive performance of our model compared with benchmark models. We report the results for both ratio prediction and weight prediction. Additionally, we provide an evaluation that excludes persistent edges to further assess the model's performance.

As mentioned in the introduction, relatively few models are capable of predicting both the formation and dissolution of links, along with their weighted values. Moreover, most models that can make such predictions are designed for small, dense networks, thereby rendering them computationally impractical for large, complex networks\footnote{The model in \cite{Zhang2022} addresses interbank networks; however, the interbank network is much smaller than real bank transfer networks, and that study focuses solely on dynamic link prediction.}. Additionally, in the dynamic link prediction literature, it is well known that because of edge persistence, simple predictions based on past link occurrences often outperform deep learning models, which makes them a strong benchmark\cite{Poursafaei2022}. These findings underscore the importance of comparing predictive performance when persistent edges are excluded. Thus, we compared our model with the following models:
\begin{itemize} 
\item \textbf{EdgeBank}: This model makes predictions by averaging all past edge weights. 
\item \textbf{EdgeBank-time-windowed}: This model uses the edge weights from the most recent month for prediction. 
\item \textbf{DLF-flat}: This is our proposed model, which uses a flat softmax. 
\item \textbf{DLF-hierarchical}: This is our proposed model, which uses a shallow and wide hierarchical softmax. 
\end{itemize}

We evaluated the models focusing on both predicting the distribution of remittance ratios and forecasting the total remittance volume. To predict the distribution of remittance ratios, we used average cross-entropy defined as follows:

\begin{equation}
    \label{equ : BCE}
    \text{bce}(R(t), \hat{R}(t)) = -\frac{1}{N}\sum_{i=1}^{N} \left( R_{i}(t) \log \hat{R}_i(t) + (1-R_i(t)) \log (1-\hat{R}_i(t)) \right),
\end{equation}

\noindent where $\hat{p}_{i}$ denotes the predicted ratio and $p{i}$ the true ratio. To evaluate the total remittance volume, we used the mean absolute error (MAE) and mean absolute percentage error (MAPE).

\begin{table}[h]
    \centering
    \caption{Results of remittance ratio predictions}
     \label{res:ratio}
    \begin{tabular}{lcc}
        \toprule
         & \textbf{Ethereum}  & \textbf{Bank} \\
        \midrule
        \textbf{EdgeBank} & 5.213 & 3.577 \\
        \textbf{EdgeBank\_tw} & 13.677 & 6.430 \\
        \textbf{DLF-Flat} & 3.694 & 3.489 \\
        \textbf{DLF-Hier} & \textbf{3.431} & \textbf{3.480} \\
        \bottomrule
    \end{tabular}
\end{table}

Table~\ref{res:ratio} presents the predictive performance for remittance ratio predictions. Our proposed model outperformed the baseline, which demonstrates its superior predictive capability. Particularly for the Ethereum dataset, it achieved high performance. Additionally, incorporating hierarchical softmax led to a slight but notable improvement in performance, which highlights the value of such adjustments, as shown by the comparison of DLF-Flat and DLF-Hier.

\begin{table}[h]
    \centering
    \caption{Results of remittance volume predictions}
    \label{res:vol}
    \begin{tabular}{lcccc}
        \toprule
        & \textbf{Eth (MAE)} & \textbf{Eth (MAPE)} & \textbf{Bank (MAE)} & \textbf{Bank (MAPE)} \\
        \midrule
        \textbf{EdgeBank} & \textbf{0.506} & 0.349 & $9.167 \times 10^{4}$ & 0.136 \\
       \textbf{EdgeBanktw} & 0.571 & 0.716 & $\mathbf{5.093 \times 10^{4}}$ & 0.351 \\
        \textbf{DLF-Flat(EB)} & 0.507 & \textbf{0.347} & $5.826 \times 10^{4}$ & 0.136 \\
        \textbf{DLF-Flat(EBtw)} & 0.603 & 1.109 & $9.170 \times 10^{4}$ & 0.385 \\
        \textbf{DLF-Flat(Pred)} & 0.509 & 21.279 & $9.097 \times 10^{4}$ & 0.133 \\
        \textbf{DLF-Hier(EB)} & 0.507 & \textbf{0.347} & $9.170 \times 10^{4}$ & 0.136 \\
        \textbf{DLF-Hier(EBtw)} & 0.601 & 1.121 & $6.829 \times 10^{4}$ & 0.382 \\
        \textbf{DLF-Hier(Pred)} & 0.509 & 21.652 & $9.098 \times 10^{4}$ & \textbf{0.132} \\
        \bottomrule
    \end{tabular}
\end{table}

\begin{figure}[h]
    \centering
    \begin{subfigure}{0.45\textwidth}
        \centering
        \includegraphics[width=\linewidth]{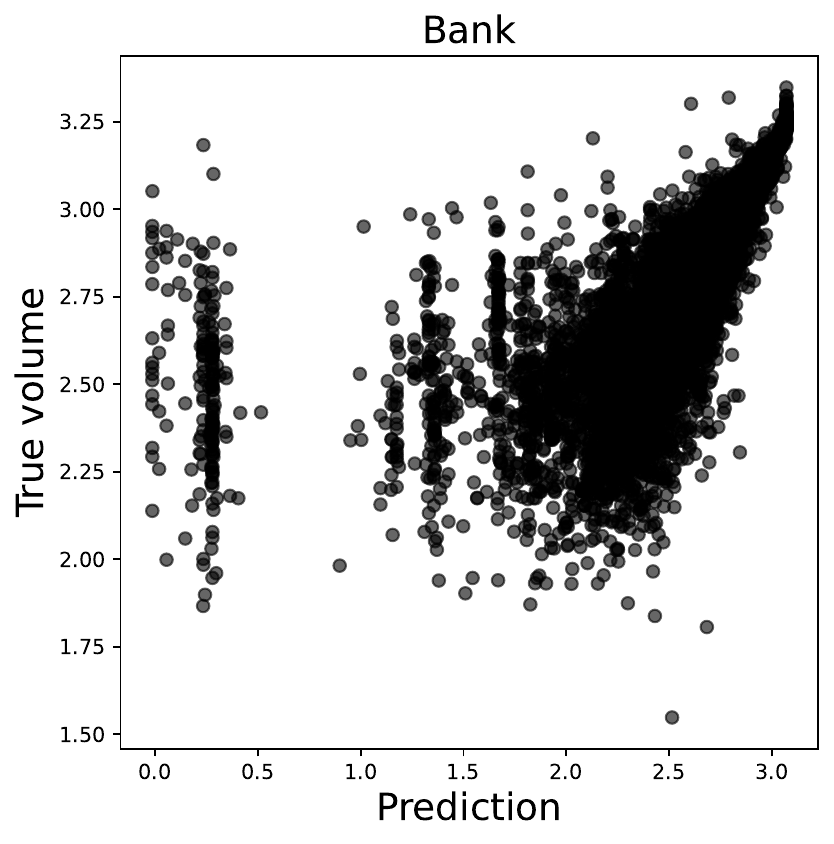}
        \caption{Bank volume}
    \end{subfigure}
    \hfill
    \begin{subfigure}{0.45\textwidth}
        \centering
        \includegraphics[width=\linewidth]{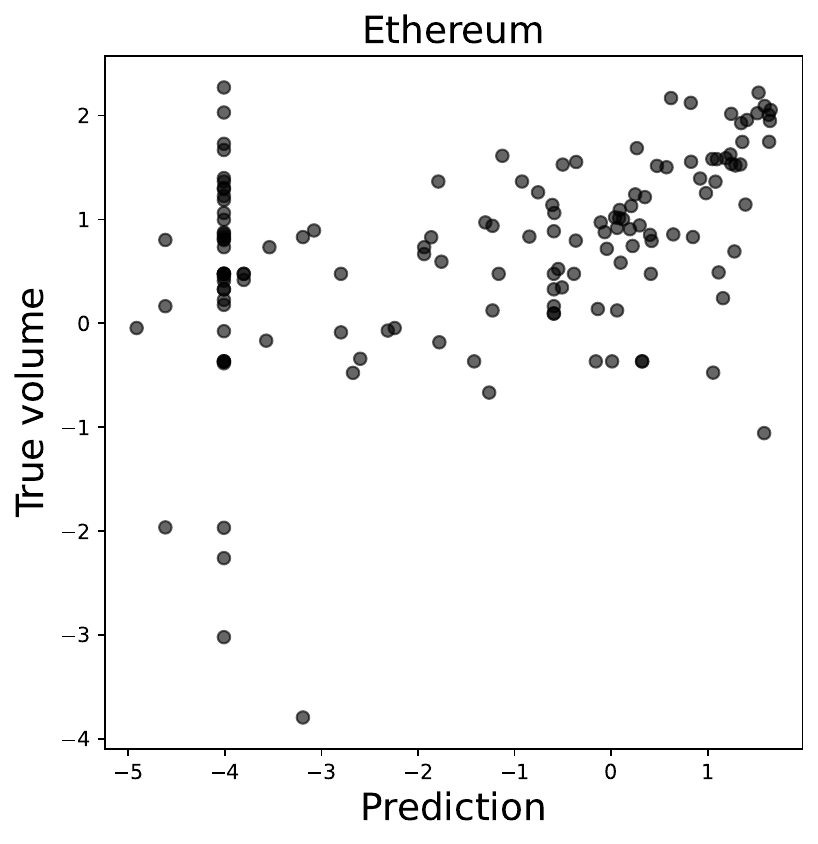}
        \caption{Ethereum volume}
    \end{subfigure}
    \caption{Comparison of bank and ethereum volumes shown in log-log plots}
    \label{fig:pred_vol}
\end{figure}

Incorporating remittance volume prediction presented more subtle challenges. As shown in Table \ref{res:vol}, although our model outperformed the baseline in some cases (e.g., Bank (MAPE)), the simple yet robust baseline surpassed our model in others. We also tested whether using past volume data within the same logical framework as EdgeBank and EdgeBank-time-windowed would improve performance. In some instances, it did (e.g., Eth (MAPE)), but in others, it did not. These mixed results highlight the inherent difficulty of predicting volume. Despite the adjustments described in the dataset section, money was only remitted occasionally for some nodes in both datasets. Additionally, fund transfers were temporarily stopped for some accounts for several months and then suddenly resumed with substantial amounts. This pattern is evident in Fig. \ref{fig:pred_vol}, where we plotted the predicted volumes for all nodes in the test data. Although there is some correlation between the predicted and actual values, many nodes show significantly lower predicted amounts because of minimal activity in the past three months; the actual values were substantially higher. In future work, we should take this characteristic into account and explore more complex models to better capture aggregate remittance volumes.

\begin{table}[h!]
\centering
\caption{Link formation and dissolution performance}
\label{tab:link_metrics}
\begin{tabular}{lcc}
\hline
\textbf{Dataset (Threshold)} & \textbf{Link Formation} & \textbf{Link Dissolution} \\
\hline
Ethereum (thres=0.0001) & 0.733 & 0.834 \\
Ethereum (thres=0.001)  & 0.769 & 0.826 \\
Bank (thres=0.0001)     & 0.650 & 0.652 \\
Bank (thres=0.001)      & 0.892 & 0.657 \\
\hline
\end{tabular}
\end{table}

We further investigated the model's performance by excluding persistent edges. We predicted link formation on all edges not included in the EdgeBank prediction. We set the label for link formation based on whether a new edge appeared in the next step, whereas we determined the prediction value using the model's probability. We predicted the link dissolution on all edges in the EdgeBank prediction, with the label indicating the occurrence of link dissolution, and set the prediction value to 1 minus the model's probability. To assess whether our model could accurately predict the formation or dissolution of edges with high probability, we tested two settings, cutting and renormalizing all edges with probabilities below $10^{-4}$ and $10^{-3}$.

Table \ref{tab:link_metrics} presents the results of our model's performance for link prediction (formation) and link dissolution, evaluated using the Area Under the Receiver Operating Characteristic curve (AUC-ROC) metric. Since the baseline models were unable to make accurate predictions, resulting in merely random performance, we excluded their results. The table clearly demonstrates that our model outperformed the random baseline of 0.5. Notably, for the Ethereum data, link dissolution prediction proved easier than link formation, which contrasts with the common assumption in the literature that link dissolution is typically more challenging\cite{Hisano2018}. Additionally, for the bank data, setting a higher threshold enhanced predictive performance, which indicates that our model accurately predicted the occurrence of edges with significant proportions.

\section{Conclusion}\label{data}
In this paper, we introduced a novel model for predicting dynamic links and flows in networks. The model combines structural embedding, transfer activity embedding, a shallow and wide hierarchical softmax tree, and a projection mechanism to address the challenge of flow prediction in complex, sparse networks. Unlike previous research on small, densely connected networks, our approach simultaneously predicts both the existence of edges and their weights. We validated our model using proprietary bank transaction data from a major bank in Japan and public cryptocurrency data. We demonstrated that our approach achieved promising results\footnote{This research was supported by JST FOREST Program (Grant Number JPMJFR216Q). We thank Edanz for editing a draft of this manuscript. }.

\bibliographystyle{spmpsci} 
\bibliography{refs} 

\end{document}